\documentclass[prl,twocolumn,floatfix,amssymb,amsmath]{revtex4}

\usepackage{amsmath}
\usepackage{graphicx}

\begin{document}

% Classification: Physical Sciences (Chemistry), Biological Sciences (Biophysics)
% \smallskip

\title{Nonlinear Relaxation Dynamics in Elastic Networks and Design Principles of Molecular Machines}
\author{Yuichi Togashi}
\altaffiliation[Present address: ]{Nanobiology Laboratories, Graduate School of Frontier Biosciences, Osaka University, 1-3 Yamadaoka, Suita, Osaka 565-0871, Japan}
\email{togashi@phys1.med.osaka-u.ac.jp}
\author{Alexander S. Mikhailov}
\email{mikhailov@fhi-berlin.mpg.de}
\affiliation{Abteilung Physikalische Chemie, Fritz-Haber-Institut der Max-Planck-Gesellschaft, Faradayweg 4-6, 14195 Berlin, Germany}

\begin{abstract}
Analyzing nonlinear conformational relaxation dynamics in elastic
networks corresponding to two classical motor proteins,
we find that they respond by well-defined internal mechanical motions
to various initial deformations and that these motions are robust against
external perturbations.
We show that this behavior is not characteristic for random
elastic networks.
However, special network architectures with such properties can be
designed by evolutionary optimization methods.
Using them, an example of an artificial elastic network,
operating as a cyclic machine powered by ligand binding, is constructed.
\end{abstract}

\maketitle

\section{Introduction}

Understanding design principles of single-molecule machines is
a major challenge.
Experimental and theoretical studies of proteins,
acting as motors \cite{Amos,Spudich,Corrie,Kitamura,Boyer},
ion pumps \cite{Gouaux,Scarborough,Toyoshima} or channels \cite{Gouaux,Perozo},
and enzymes \cite{Blumenfeld,Adams,Reed,Hess,Rigler},
show that their operation involves
functional conformational motions (see \cite{Gerstein}).
Such motions are slow and cannot therefore be reproduced by
full molecular dynamics simulations.
Within the last decade, approximate descriptions based on
elastic network models of proteins have been
developed \cite{Tirion,Bahar1997,Haliloglu,Tama,Liao,Bahar}.
In this approach,
structural elements of a protein are viewed as
identical point particles, with two particles connected by an elastic
string if the respective elements lie close enough in the native state of
the considered protein.
Thus, a network of elastic connections corresponding
to a protein is constructed.
So far, the attention has been focused
on linear dynamics of elastic networks, characterized in terms of their
normal vibrational modes.
It has been found that
ligand-induced conformational changes in many proteins agree with the patterns of
atomic displacements in their slowest vibrational
modes \cite{Karplus,Levitt,Zheng,Li} (see also \cite{Baharbook,Yang}),
even though
nonlinear elastic effects must become important for large deviations from
the equilibrium \cite{Ma,Miyashita}.
The focus of this article is on nonlinear relaxation phenomena in elastic networks
seen as complex dynamical systems.

Generally, a machine is a mechanical device
that performs ordered internal motions which are robust
against external perturbations.
In machines representing single molecules,
energy is typically supplied in discrete portions,
through individual reaction events.
Therefore, their cycles consist of the
processes of conformational relaxation that follow after energetic excitations.
For a robust machine operation, special nonlinear
relaxation dynamics is required.
We expect that, starting from a broad range of initial deformations,
such dynamical systems would return to the same final equilibrium state.
Moreover, the relaxation would proceed along a well-defined trajectory
(or a low-dimensional manifold),
rapidly approached starting from different initial states
and robust against external perturbations.
These attractive relaxation trajectories would define internal
mechanical motions of the machine inside its operation cycle.

This special conformational relaxation dynamics has been confirmed in
our study of the elastic networks
of two protein motors (F$_{1}$-ATPase and myosin).
On the other hand,
our control investigation of random elastic networks has shown
that relaxation patterns in random elastic networks are typically
complex and qualitatively different from those of protein motors.
Actual proteins with specific architectures allowing robust machine
operation may have developed through a natural biological evolution,
with the selection favoring such special dynamical properties.
In a model study,
we have demonstrated that artificial elastic network architectures
possessing machine-like properties can be designed
by running an evolutionary computer optimization process.
Finally, an example of an artificially designed elastic network
that operates like a machine powered
by ligand binding has been constructed.

\subsection*{Elastic network models}

\begin{figure*}
\begin{center}
\includegraphics[height=47mm]{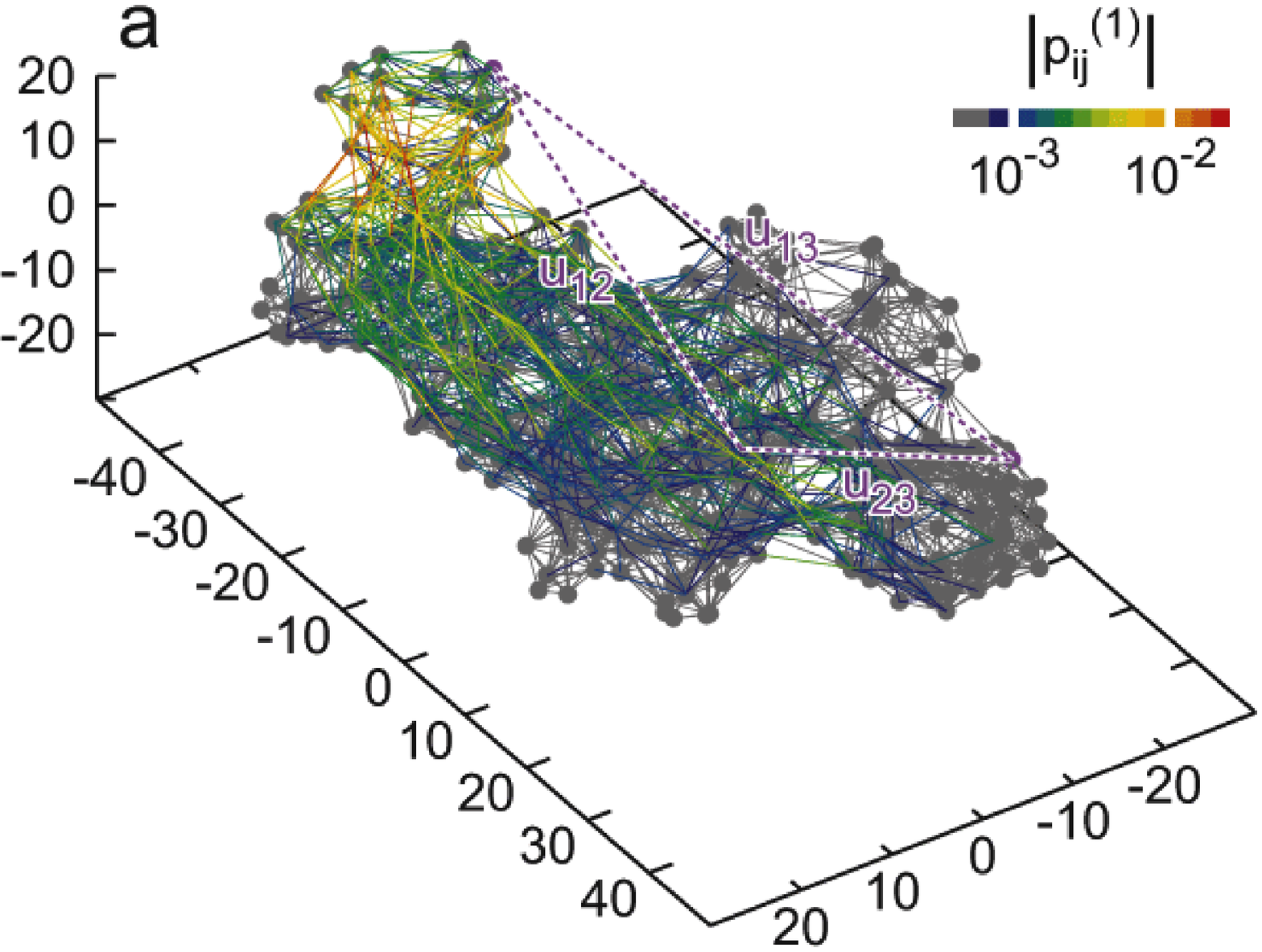}
\includegraphics[height=47mm]{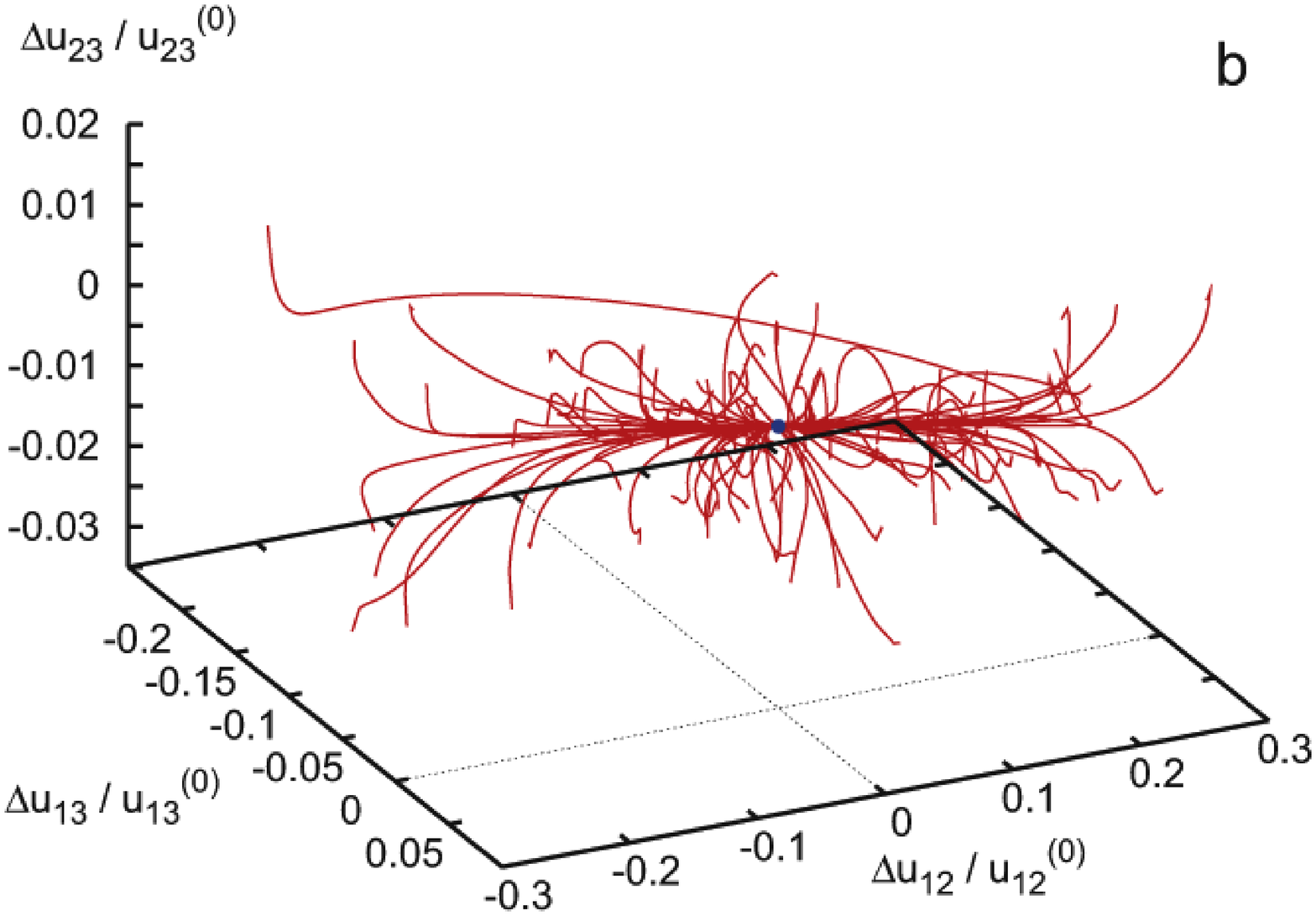}
\end{center}
\caption{Elastic network of the single $\beta$-subunit of
the molecular motor F$_{1}$-ATPase (a) and the
set of relaxation trajectories for this network (b). Links are colored
according to their relative deformations
$\left\vert p_{ij}^{(1)} \right\vert$ in the motion corresponding to
the slowest normal mode. Each of 100
trajectories starts from a different initial conformation (see Methods). In
this case, all the trajectories converge to the original equilibrium state
(blue dot). The labels ($1$,$2$,$3$) are attached at Ile390, Arg191 and
Gly54, respectively.}
\label{fig1}
\end{figure*}

The considered elastic networks consist of a set of $N$ identical material
particles (nodes) connected by identical elastic strings (links).
A network is specified by indicating equilibrium positions of all particles.
Two particles are connected by a string if the equilibrium
distance between them is sufficiently small. The elastic forces, acting on
the particles, obey Hooke's law and depend only on the change in the
distances between them. In the overdamped limit \cite{Haliloglu}, the
velocity of a particle is proportional to the sum of elastic forces applied
to it. If $\mathbf{R}_{i}^{(0)}$ are equilibrium positions of the particles
and $\mathbf{R}_{i}(t)$ are their actual coordinates, the dynamics is
described by equations
\begin{equation}
\overset{.}{\mathbf{R}}_{i}=-\sum_{j=1}^{N}A_{ij}\frac{\mathbf{R}_{i}-\mathbf{R}_{j}}{\left\vert \mathbf{R}_{i}-\mathbf{R}_{j}\right\vert }\left(\left\vert \mathbf{R}_{i}-\mathbf{R}_{j}\right\vert -\left\vert \mathbf{R}_{i}^{(0)}-\mathbf{R}_{j}^{(0)}\right\vert \right)
\label{nonlinear}
\end{equation}
where $A$ is the adjacency matrix, with the elements $A_{ij}=1$,
if $\left\vert \mathbf{R}_{i}^{(0)}-\mathbf{R}_{j}^{(0)}\right\vert <l_{0}$,
and $A_{ij}=0$ otherwise.
The dependence on the stiffness constant of the strings
and the viscous friction coefficient of the particles is removed
here by an appropriate rescaling of time.

The dynamics of elastic networks is nonlinear, because distances
$\left\vert \mathbf{R}_{i}-\mathbf{R}_{j} \right\vert$ are nonlinear functions of the
coordinates $\mathbf{R}_{i}$ and $\mathbf{R}_{j}$. Close to the equilibrium,
equations of motion can however be linearized, yielding
\begin{equation}
\overset{.}{\mathbf{r}}_{i}=-\sum_{j=1}^{N}A_{ij}\frac{\mathbf{R}_{i}^{(0)}-\mathbf{R}_{j}^{(0)}}{\left\vert \mathbf{R}_{i}^{(0)}-\mathbf{R}_{j}^{(0)}\right\vert ^{2}}\left[ \left( \mathbf{R}_{i}^{(0)}-\mathbf{R}_{j}^{(0)}\right) \cdot \left( \mathbf{r}_{i}-\mathbf{r}_{j}\right) \right]
\label{linear}
\end{equation}
for small deviations $\mathbf{r}_{i}=\mathbf{R}_{i}-\mathbf{R}_{i}^{(0)}$.
These equations can be written as
$\overset{.}{\mathbf{r}}_{i}=-\sum_{j}\mathbf{\Lambda }_{ij}\mathbf{r}_{j}$,
where $\mathbf{\Lambda }$ is a $3N\times 3N$ linearization matrix.
In the linear approximation, relaxation
motion is described by a sum of independent exponentially decaying normal modes
\begin{equation}
\mathbf{r}_{i}(t)=\sum_{\alpha }X_{\alpha }\mathbf{e}_{i}^{(\alpha )};\ X_{\alpha}=k_{\alpha} \exp \left( -\lambda_{\alpha}t \right)
\label{nmdecay}
\end{equation}
with $\lambda_{\alpha}$ and $\mathbf{e}_{i}^{(\alpha)}$ representing
nonzero eigenvalues and the respective eigenvectors of the matrix $\mathbf{\Lambda}$.
It should be noted that the same eigenvalues determine vibration
frequencies of the network,
$\omega_{\alpha} \sim \sqrt{\lambda_{\alpha}}$,
and the vibrational normal modes are the same. Long-time relaxation is
dominated by soft modes with small eigenvalues.

\section{Results}

\subsection*{Two motor proteins}

\begin{figure*}
\begin{center}
\includegraphics[height=84mm]{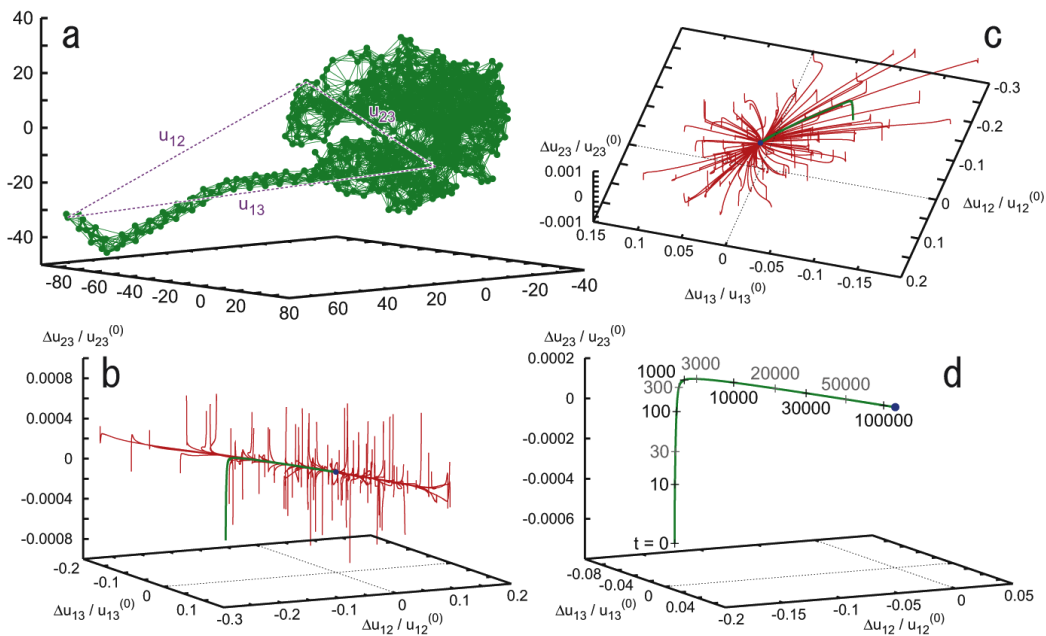}
\end{center}
\caption{Elastic network of the single heavy chain of myosin (a)
and the set of relaxation trajectories for this network (b,c,d).
Panels (b) and (c) are viewed from different angles;
each of 100 trajectories starts from a different
initial conformation (see Methods).
All the trajectories converge to the original equilibrium state (blue dot).
Panel (d) shows a trajectory (shown by green curve in (b) and (c)) with
labels of passage time.
The labels ($1$,$2$,$3$) are attached at Leu836, Asp63 and Glu370, respectively.}
\label{fig2}
\end{figure*}

\begin{figure}
\begin{center}
\includegraphics[width=80mm]{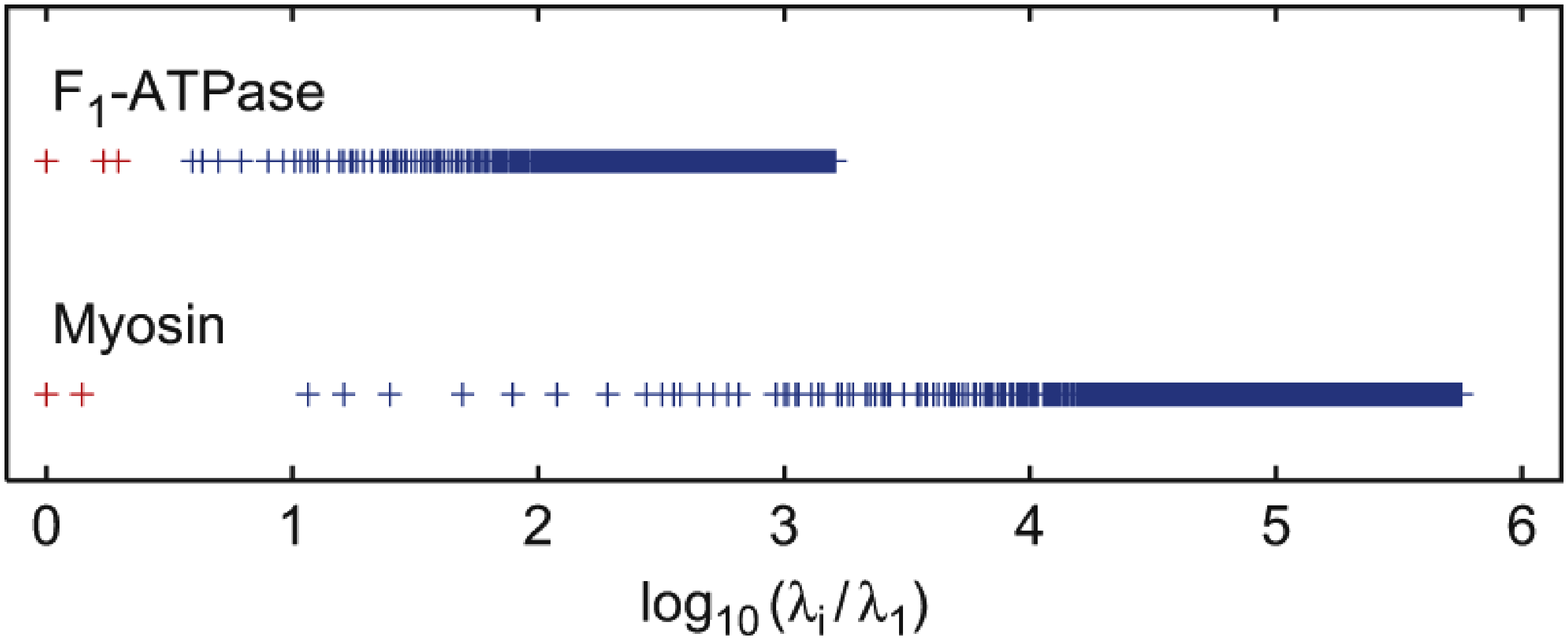}
\end{center}
\caption{Eigenvalue spectra for the networks of F$_{1}$-ATPase and myosin,
normalized to the lowest nonzero eigenvalue $\lambda_{1}$.
$\lambda_{1} = 1.09 \times 10^{-2}$ for F$_{1}$-ATPase
and $2.81 \times 10^{-5}$ for myosin.
}
\label{fig3}
\end{figure}

As an example,
Fig. 1 shows the elastic network of the single $\beta$-subunit of the
molecular motor F$_{1}$-ATPase (Protein Data Bank ID: 1H8H, chain E) \cite{1H8H}.
Each node corresponds to a residue in this protein (the total
number of nodes is $N=466$). In Fig. 1b, a set of conformational relaxation
trajectories, obtained by numerical integration of the nonlinear elastic
model (see Methods, Eq. (1)) of this macromolecule, is displayed.
To track conformational motions,
three network nodes ($1$, $2$ and $3$) were chosen
and pair distances $u_{12}$, $u_{13}$ and $u_{23}$ were determined during the
relaxation process. Thus, each conformational motion was represented by a
trajectory in a three-dimensional space, where coordinates were normalized
deviations $\Delta u_{ij}/u_{ij}^{(0)}$ from the equilibrium pair distances
$u_{ij}^{(0)} = \left\vert \mathbf{R}_{i}^{(0)}-\mathbf{R}_{j}^{(0)} \right\vert$ ($u_{ij} = \left\vert \mathbf{R}_{i}-\mathbf{R}_{j} \right\vert$ and
$\Delta u_{ij} = u_{ij}-u_{ij}^{(0)}$).
Each trajectory begins from a different
initial conformation obtained by applying random static forces to all
network nodes (see Methods).
Trajectories starting from
various initial conditions soon converge to a well-defined relaxation path
leading to the equilibrium state. This path corresponds to a slow motion of
the network group, including label 1, with respect to the rest of the
molecule. The protein is soft along such a path: by applying static forces
of the same magnitude, one can stretch it by 30\% along the path, as
compared to the length changes of only a few percent when the forces were
applied in other directions.

Relaxation behavior in the nonlinear elastic network ($N=793$) of
another classical molecular motor, myosin (single heavy chain;
PDB ID: 1KK8, chain A) \cite{1KK8}, is displayed in Fig. 2.
In contrast to the $\beta$-subunit of F$_{1}$-ATPase,
this elastic network possesses an attractive
two-dimensional manifold (a plane).
The network is extremely stiff for
deformations in the directions orthogonal to this plane. By applying static
forces of the same magnitude, one can only induce relative deformations of
about $10^{-3}$ along such orthogonal directions, as compared to the
relative deformations of the order of $10^{-1}$ for the directions within
the plane.
To characterize the temporal course of relaxation,
one relaxation trajectory (displayed by green color in Fig. 2 (b,c)) and
subsequent positions at different time moments along the trajectory
are indicated in Fig. 2d.
The trajectory rapidly reaches the plane
and then the relaxation motion becomes much slower
(with the characteristic time scales larger by a factor $10$ to $100$).
A similar behavior is characteristic for other relaxation trajectories,
starting from different initial conditions.
All recorded trajectories returned to the equilibrium
state and no metastable states were encountered starting from the chosen
initial conditions.

Nonlinear effects were essential in the relaxation dynamics
starting from large arbitrary initial deformations considered here.
Remarkably, the observed relaxation patterns are nonetheless qualitatively
in agreement with the normal mode analysis.
Both motors possess a group of soft modes
separated by a gap from the rest of the spectrum
(eigenvalue spectra of elastic networks of these proteins are shown
in Fig. 3).
The two soft modes of myosin define the attractive plane seen in the relaxation
pattern of its elastic network (Fig. 2). The elastic network of
F$_{1}$-ATPase has three soft modes. However, one of the soft modes has the
relaxation rate which is smaller than the other two modes.
Therefore, the pattern of relaxation trajectories looks here like
a thick one-dimensional bundle.
Specific ligand-induced conformational changes in F$_{1}$-ATPase and myosin
were previously shown to have strong overlaps with patterns of deformation
in slow vibrational modes \cite{Zheng}.

\subsection*{Random and designed elastic networks}

\begin{figure}
\begin{center}
\includegraphics[height=42mm]{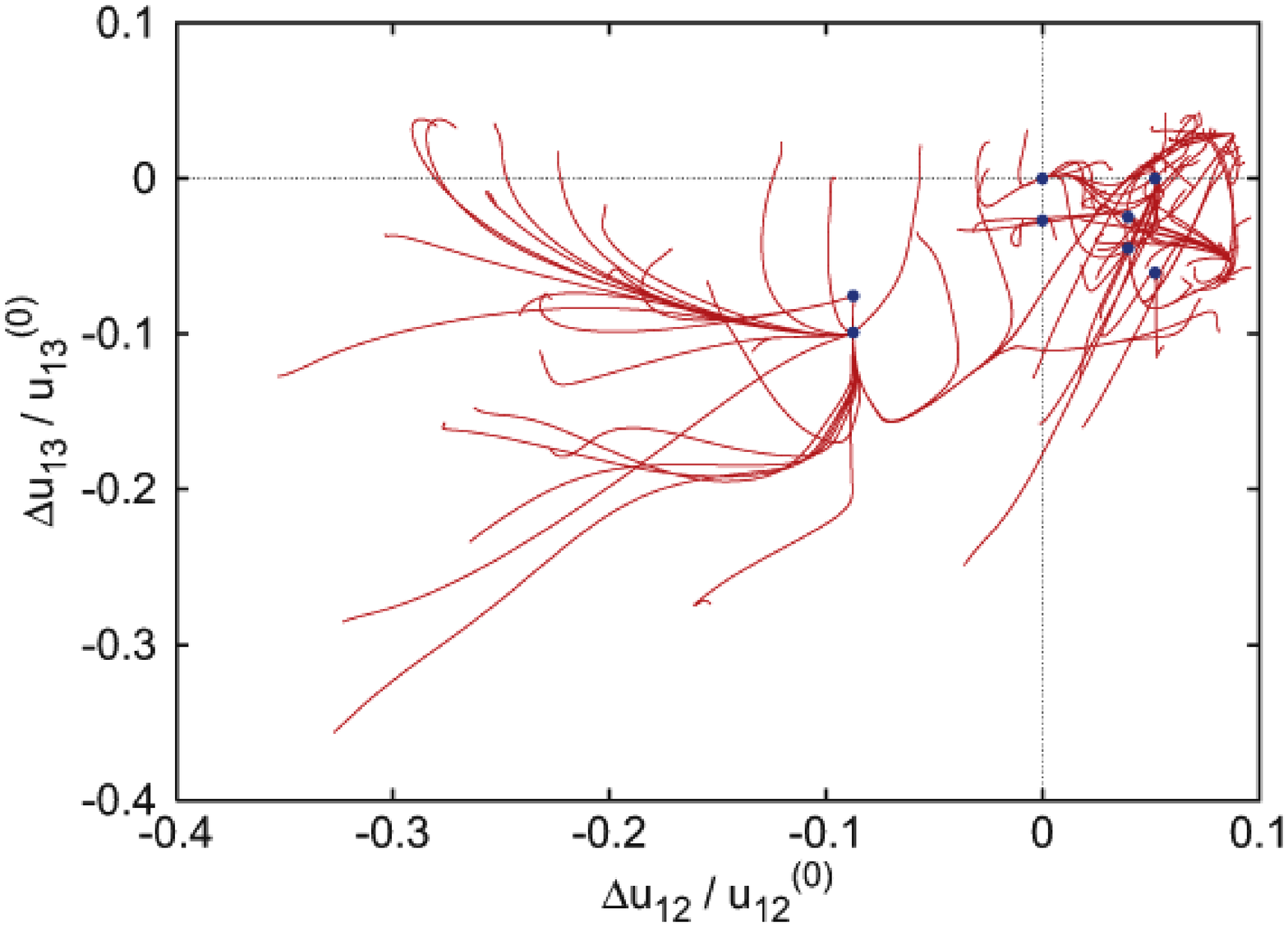}
\includegraphics[height=42mm]{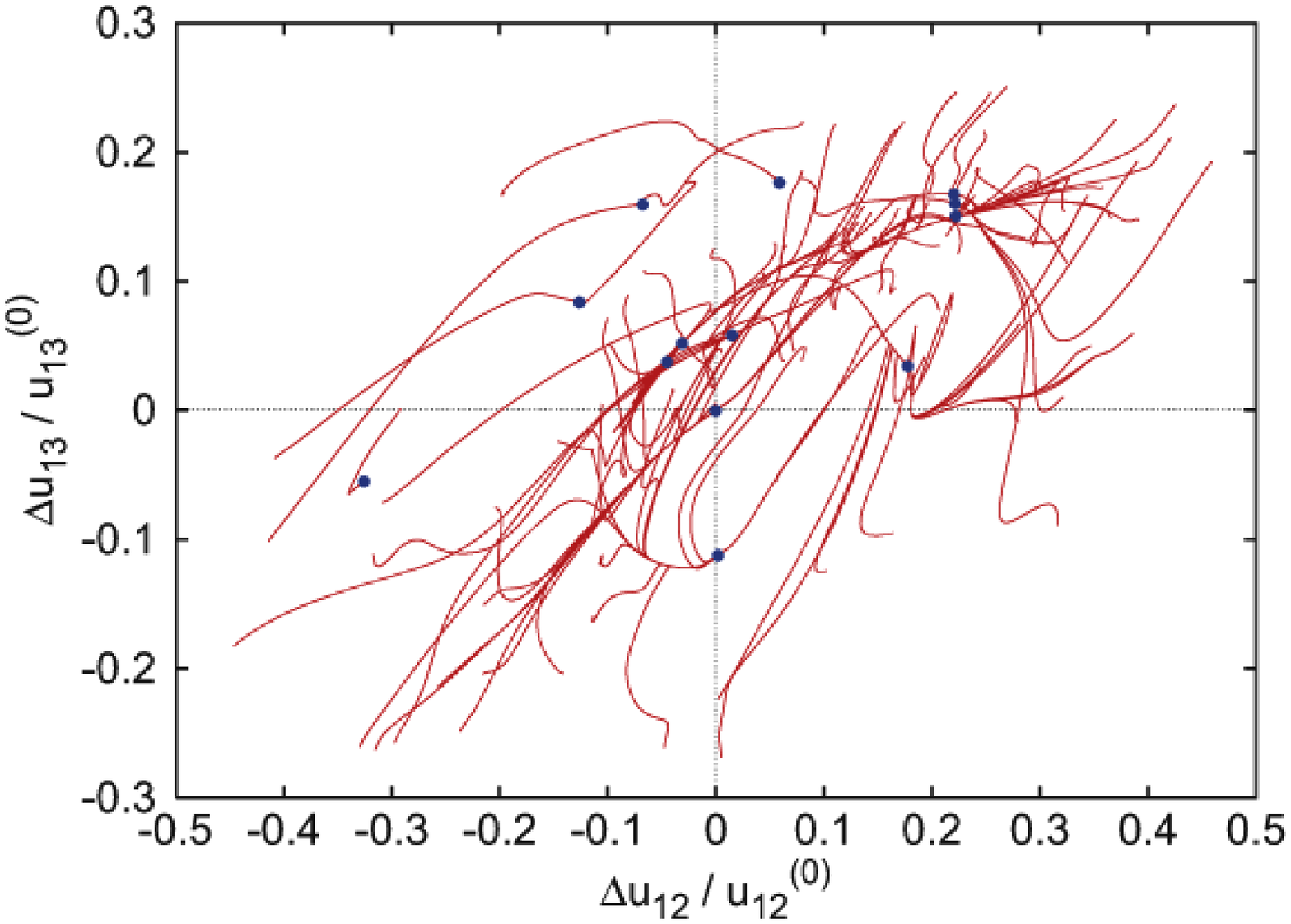}
\end{center}
\caption{Relaxation trajectories for two random elastic networks, in the
plane ($\Delta u_{12}/u_{12}^{(0)}$, $\Delta u_{13}/u_{13}^{(0)}$) of
normalized distance deviations. Each of 100 trajectories starts from a
different initial conformation.
Blue dots indicate stationary states reached;
the original equilibrium state is $\Delta u_{12} = \Delta u_{13} = 0$.
These networks have no internal rotational modes.
}
\label{fig4}
\end{figure}

A control study of nonlinear relaxation phenomena in random elastic networks
was performed.
Such networks were obtained by taking a relatively short
chain of $N=64$ and randomly folding it in absence of energetic interactions
(see Methods).
After that, all particles separated by short enough distances
were connected by identical elastic links.
Figure 4 shows relaxation
patterns for two typical random elastic networks (using the same procedure
for generation of initial conditions and for tracking of conformational
relaxation as in Figs. 1 and 2).
Relaxation patterns in random networks are
clearly different from those of the considered motor proteins.
Random networks possess many (meta)stable stationary states, with relaxation
trajectories often ending in one of them instead of going back to the
equilibrium conformation.
The linear normal mode description holds in such
networks only in close proximity of the equilibrium state.

Thus, elastic networks of motor proteins are special.
Their equilibrium conformation has a big attraction basin.
Starting from an arbitrary initial deformation,
relaxation dynamics is soon reduced to a low-dimensional
attractive manifold.
Within its large part,
the dynamics is approximately linear and determined
by a few soft modes.
Proteins with such special dynamical properties,
essential for their functions,
may have emerged through a biological evolution.
Below, we show that artificial elastic networks
with similar properties can be constructed
by running an evolutionary optimization process
based on a variant of the Metropolis algorithm.

The evolutionary optimization technique is described in Methods.
For each network, spectral gap
$g = \log_{10} \left( \lambda_{2}/\lambda_{1} \right)$ is
defined as the logarithm of the ratio between the
relaxation rates $\lambda_{2}$ and $\lambda_{1}$ of its two slowest normal
modes. If a substantial gap is present, the slowest mode has the relaxation
rate which is much smaller than that of the other modes; therefore, the
long-time relaxation in the linear regime would be dominated by this soft
normal mode. The employed evolutionary optimization process maximizes the
spectral gap $g$ of the evolving networks. Beginning with an initial random
network, we applied structural mutations and determined the difference
$\Delta g = g^{\prime}-g$ of the gaps before and after a mutation. If the gap
was increased ($\Delta g>0$), the mutation was always accepted.
If $\Delta g < 0$, the mutation was accepted with the probability
$P = \exp \left( \Delta g / \theta \right)$ where $\theta$ is the effective optimization
temperature. This procedure was applied iteratively, generating an evolution
started from an initial network \footnote{
The described evolutionary optimization algorithm allows to construct
only the networks with a single soft mode.
It can be, however, easily modified for the design of networks with
two or more soft modes, separated by a gap from the rest of the spectrum.}.

\begin{figure*}
\begin{center}
\includegraphics[height=42mm]{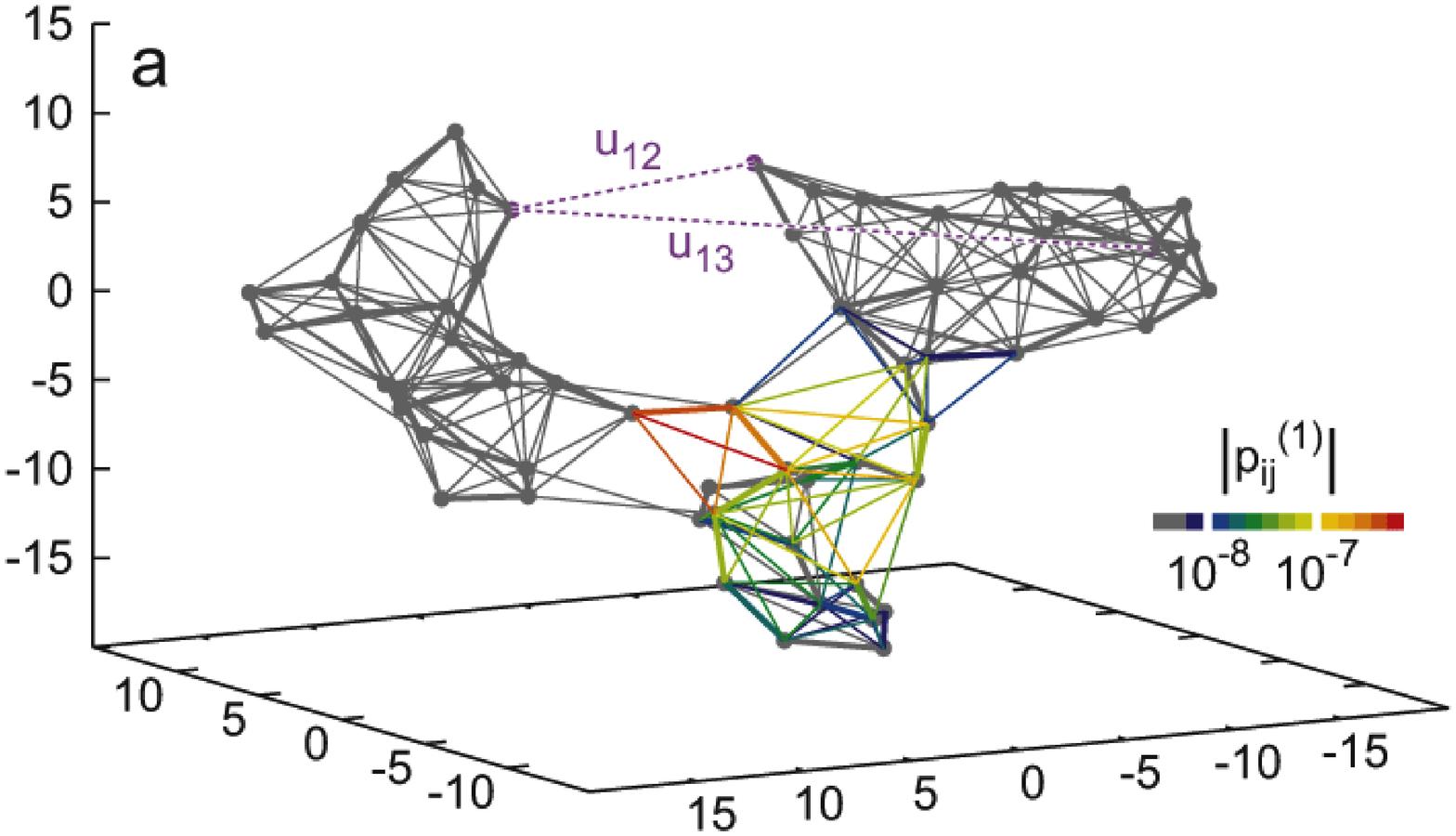}
\includegraphics[height=42mm]{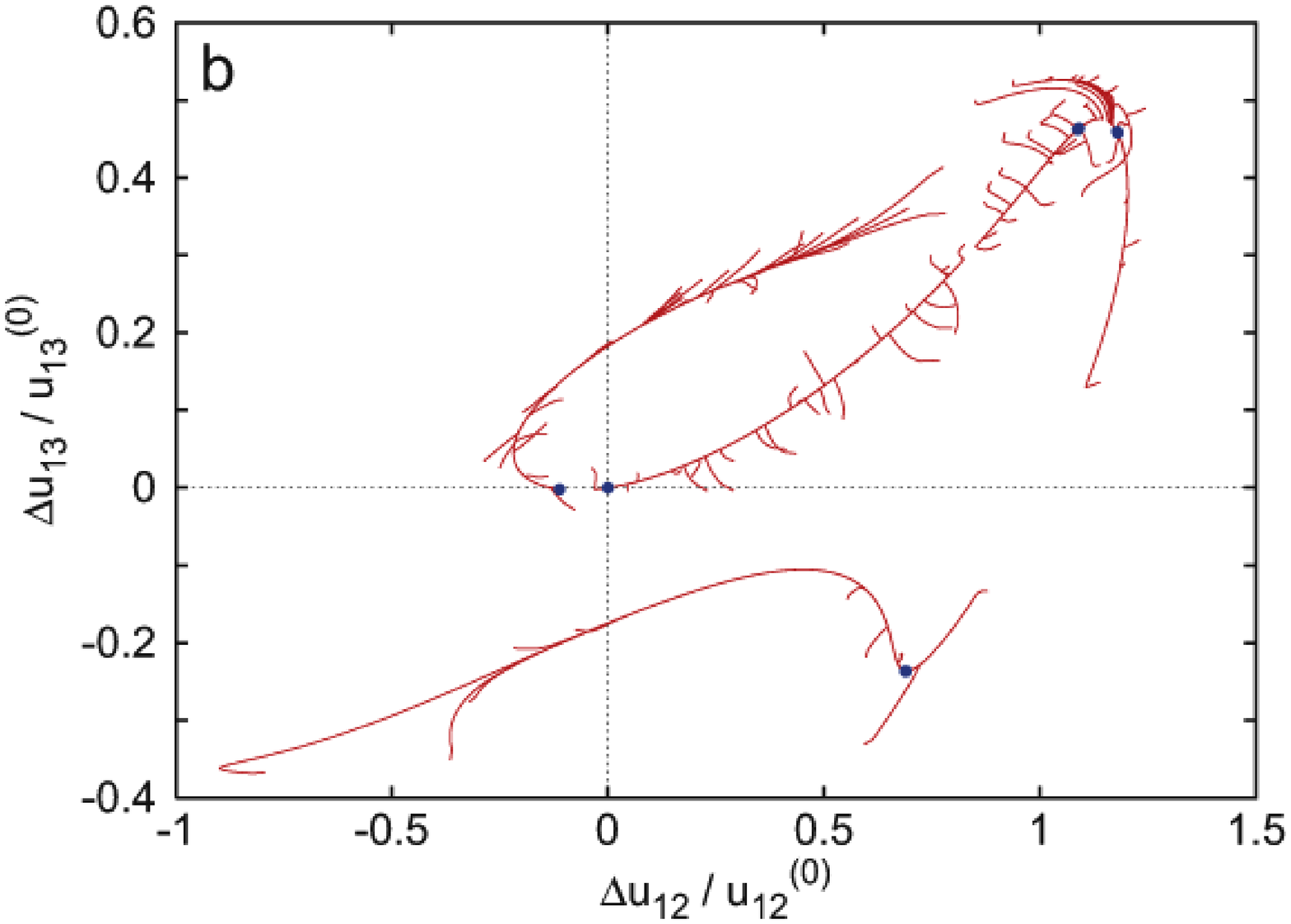}
\includegraphics[height=42mm]{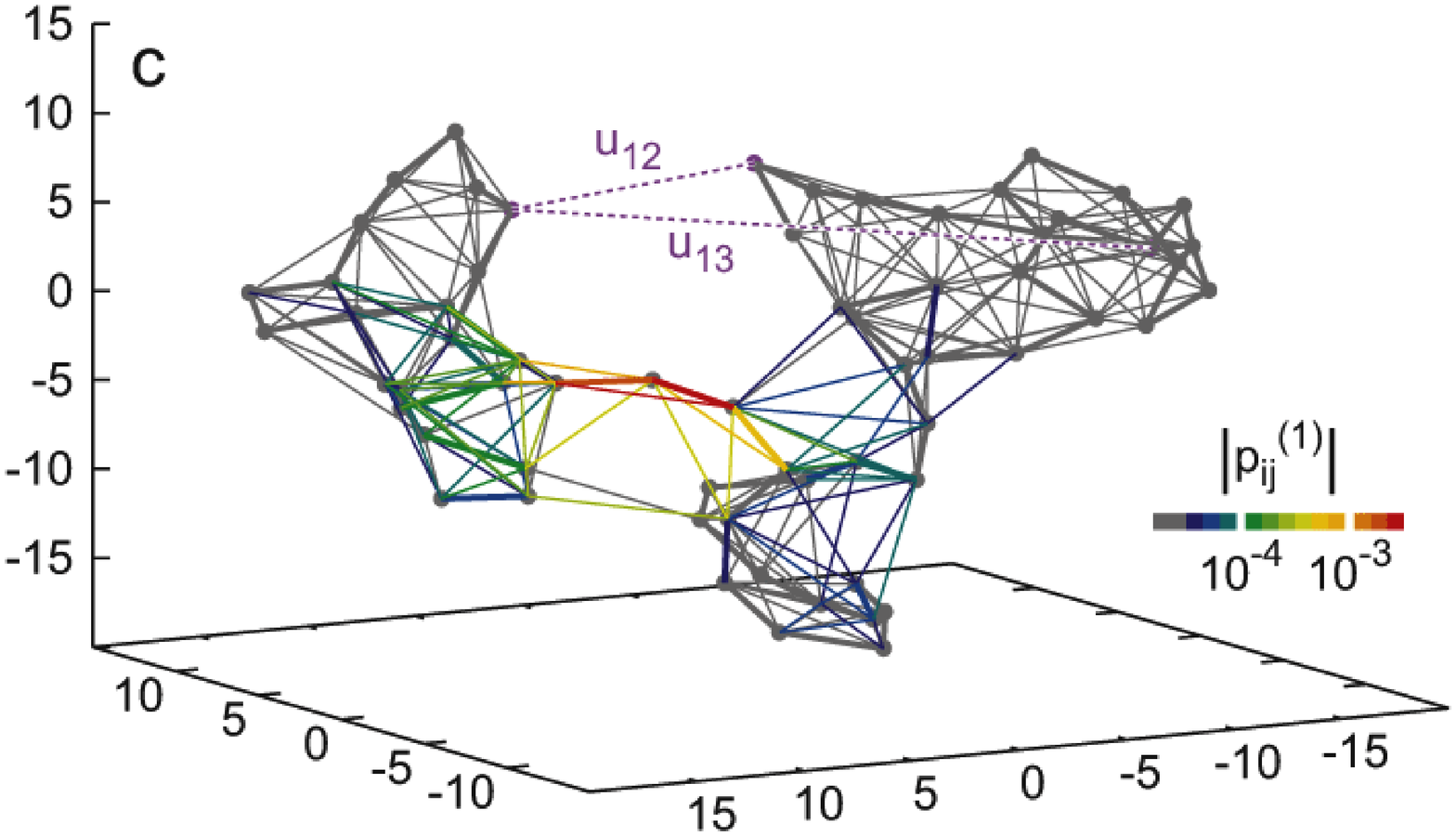}
\includegraphics[height=42mm]{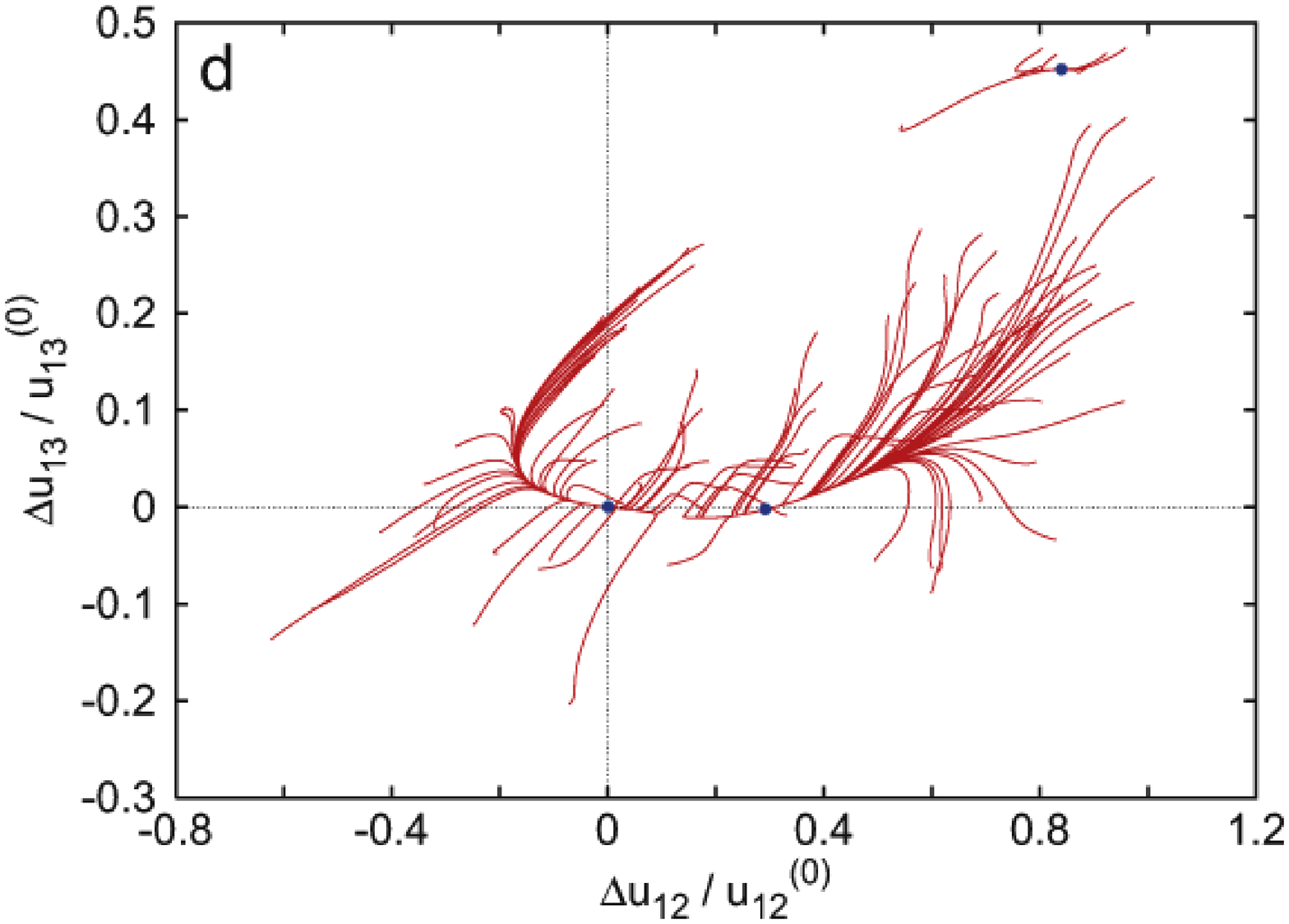}
\end{center}
\caption{Examples of designed elastic networks. (a) A typical network with a
large gap ($g = 9.53$) and (b) its set of 100 relaxation trajectories; (c) a
network with a smaller gap ($g = 1.22$) obtained by reverse evolution from
the network in panel (a), and (d) its set of 100 relaxation trajectories.
Network links are colored according to their relative deformations
$\left\vert p_{ij}^{(1)} \right\vert$ in the motion corresponding to the
slowest normal mode; thick lines indicate the backbone chain. Each
trajectory starts from a different initial conformation; blue dots indicate
stationary states reached.}
\label{fig5}
\end{figure*}

Networks with soft normal modes and large spectral gaps were thus
constructed. A typical network with a large gap and its relaxation pattern
are shown in Fig. 5 (a,b) (for other examples, see Supplementary Fig. 1).
The presence of a
large gap has a strong effect on the nonlinear relaxation properties in such
systems. They possess well-defined long paths with slow conformational
motion leading to the equilibrium state. There are only a few (meta)stable
states and these states usually lie on an attractive relaxation path, so
that a small barrier is encountered when moving along it. The opposite
behavior with complex relaxation patterns and a high number of (meta)stable
conformations was found for a set of ``failed'' networks where gaps were
small and could not be
significantly increased through the evolution (see Supplementary Fig. 2).
Our analysis shows that the designed networks can be viewed as consisting
of rigid blocks connected by soft joints;
they are able to perform some large conformational changes
accompanied by only small local deformations
(see Supplementary Fig. 3).

Spectral gaps of the networks, designed by using this optimization method,
are much larger than those characteristic for motor proteins
(cf. Fig. 3).
To improve the agreement, additional ``reverse''
evolution was subsequently applied to the designed networks,
with the selection pressure aimed to decrease the gap (see Methods).
While the gap was rapidly reduced,
the global relaxation pattern was changing only much more slowly
with structural mutations and retained characteristic features of the
networks with large gaps. Figure 5c displays the network, obtained by
applying such reverse evolution (with only 5 subsequent mutations) to the
original network shown in Fig. 5a. Although the spectral gap has been
reduced from $9.53$ to $1.22$, the principal structure of the network is
retained, with the mutations mostly affecting only the hinge region. The
relaxation pattern of the constructed network (Fig. 5d) reveals an
attractive path leading to the equilibrium state (with another stable state
lying on it).
Remarkably, the linear normal mode
description of relaxation dynamics holds in such networks within a much larger
domain around the equilibrium state.
We conclude that, by running an evolutionary optimization process,
artificial elastic networks approaching conformational relaxation
properties of real motor proteins can be constructed.

\subsection*{An artificial machine network}

\begin{figure*}
\begin{center}
\includegraphics[width=90mm]{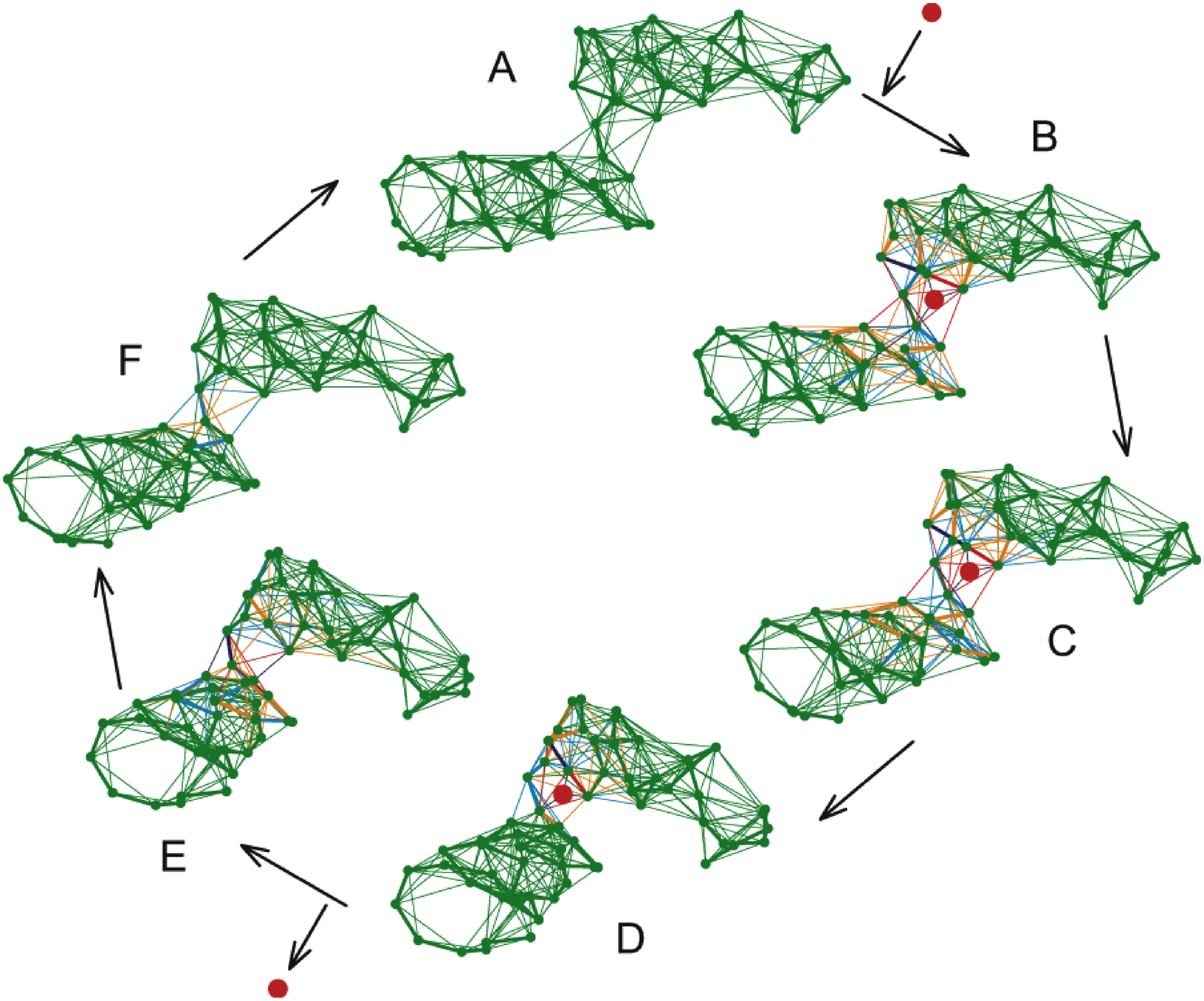}\vspace{4mm}\\
\includegraphics[height=42mm]{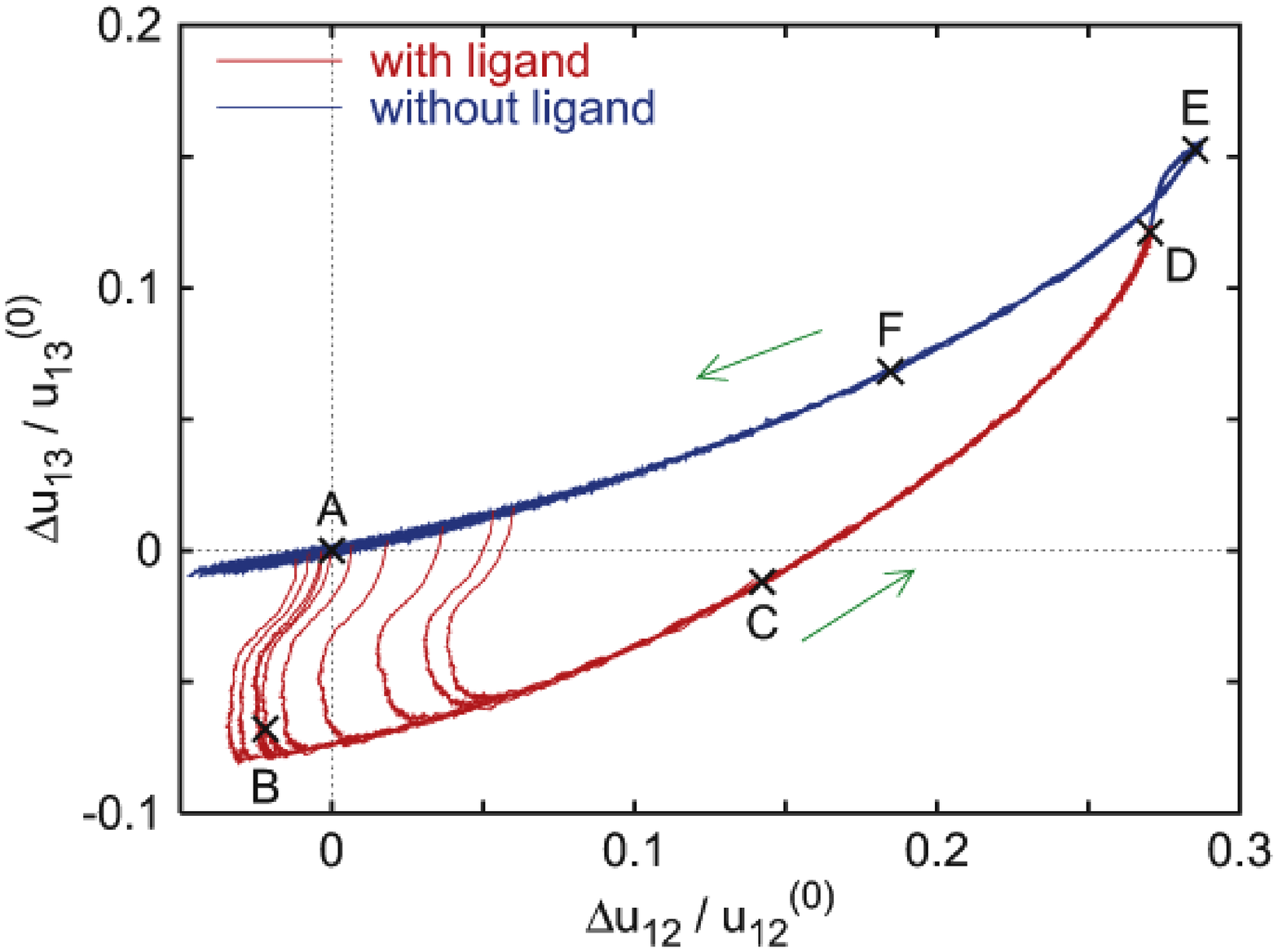}
\includegraphics[height=42mm]{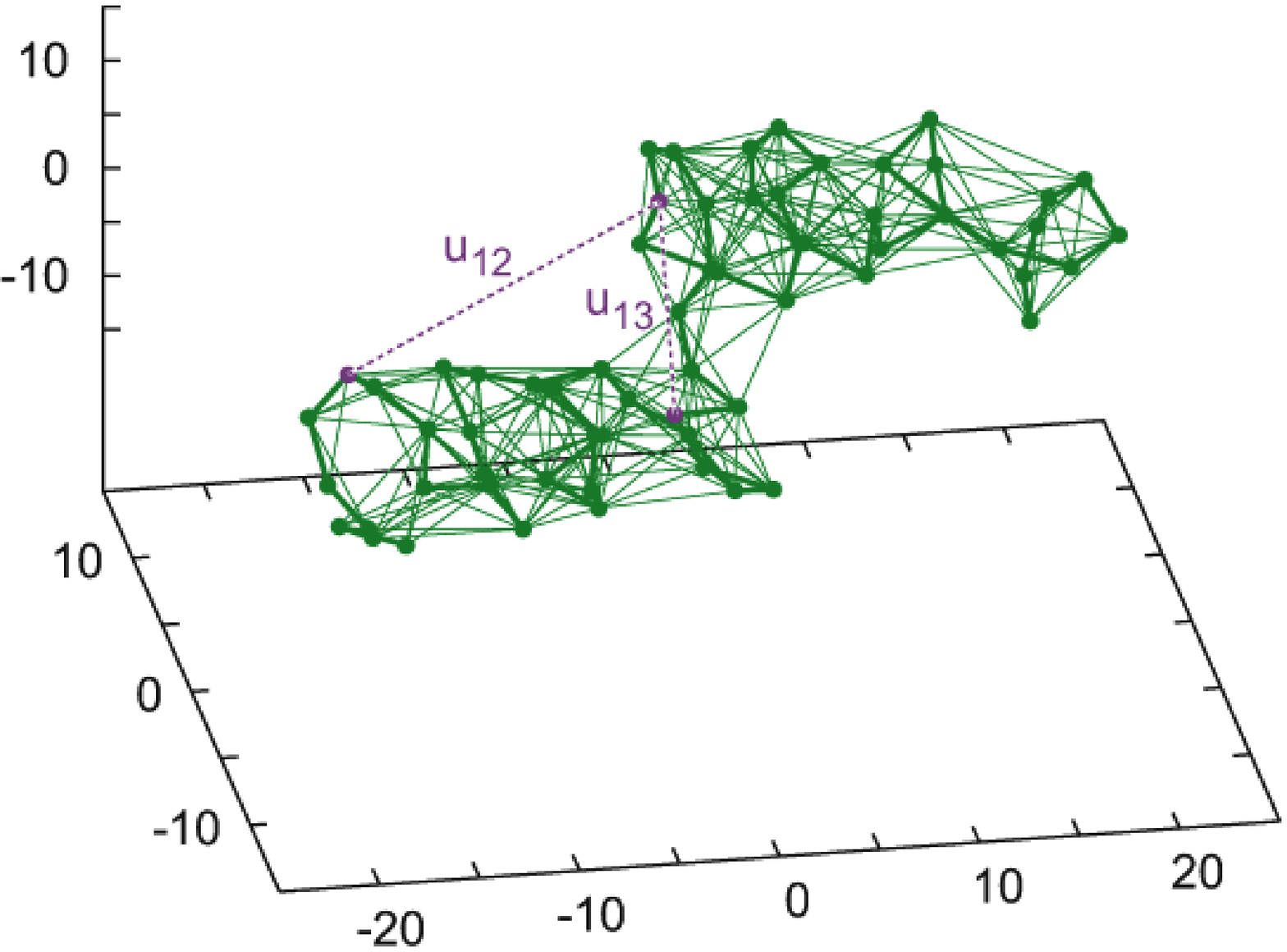}
\end{center}
\caption{The prototype of a molecular machine: an elastic network performing
regular cyclic hinge motions caused by repeated binding and detachment of
ligand particles. A series of snapshots (A--F), corresponding to different
cycle phases, is displayed. In the snapshots, the stretched links are shown
in light blue ($0.01 < \Delta u_{ij} \leq 0.1$) and
in blue ($0.1 < \Delta u_{ij}$),
the compressed links in orange ($-0.1 \leq \Delta u_{ij} < -0.01$)
and in red ($\Delta u_{ij} < -0.1$). The ligand is shown as a red dot. The
lower left panel shows trajectories of the network for $10$ subsequent
cycles in the plane ($\Delta u_{12}$,$\Delta u_{13}$) of distance deviations
between the three labeled nodes, indicated in the right panel. Dots A to F
along the trajectory correspond to the network conformations shown in the
snapshots A to F above.}
\label{fig6}
\end{figure*}

Using such designed networks, we proceed to construct nonlinear elastic
systems which can be viewed as prototypes of a machine powered by ligand
binding. The network shown in Fig. 6 performs cyclic hinge motions, caused
by binding and detachment of a ligand. To obtain it, an initial network with
two distinct parts (dense clusters), which were only loosely connected, was
prepared. This initial network was characterized by a small spectral gap.
By running evolutionary optimization, the architecture of the network was
modified, so that it developed a large spectral gap while maintaining
its special structure. The equilibrium conformation of the finally obtained
network is shown in the lower right panel in Fig. 6. The cycle began with
binding of a ligand (snapshot A, $t=0$). The ligand was an additional
particle that could form elastic connections with the existing network nodes.
Binding of a ligand was modeled by placing the particle at a location chosen
in the hinge region and creating elastic links with the three near nodes
(for details, see Methods).
When such new links were introduced, they were in a deformed
(i.e., stretched) state and a certain amount of energy was thus
supplied to the system.
Therefore, the network-ligand complex was \emph{not}
initially in the equilibrium state and conformational relaxation to the
equilibrium had to occur.
Within a short time, deformations spread over a
group of near elastic links (B, $t=200$), producing a small change in the
network configuration. After that, a slow hinge motion of the network-ligand
complex took place (C, $t=4000$) until the equilibrium state of the complex
was reached (D, $t=20000$).
In this state, the ligand was removed.
At this moment,
the elastic network was in a configuration far from equilibrium and
its relaxation set on. Within a short time, the network reached the path
(E, $t=20200$) along which its slow ordered relaxation proceeded (F, $t=30000$),
until the equilibrium conformation (A) was reached. Then, a new ligand could
bind at the network, and the cycle was repeated.
For visualization of conformational motions inside the network cycle,
see Supplementary Video 1.

To monitor conformational motions, three labels were attached to the network
and distances $u_{12}$ and $u_{13}$ between them were recorded. The lower
left panel in Fig. 6 shows trajectories in the plane
($\Delta u_{12}$,$\Delta u_{13}$)
which correspond to the operation of this prototype machine
in the presence of noise and with stochastic binding of ligands. As seen in
Fig. 6, the trajectories of the forward and back hinge motions are
different, but both of them are well defined. The applied noise (see
Methods) is only weakly perturbing them. This means that both the free
network and the network-ligand complex have narrow valleys with steep walls,
leading to their respective equilibrium states. Binding of a ligand is
possible in a relatively broad interval of conformations near the
equilibrium and therefore there is a dispersion in the transitions from the
upper to the lower branch of the cycle.
The operation of this machine is
essentially nonlinear and unharmonic effects are strong.
Characteristic rates of slow relaxation motions in this network differ
by a factor between $10$ and $1000$ from the rate of the fast conformational
transition following the ligand binding.

\section{Discussion}

In conclusion, we have shown that motor proteins possess
unique dynamical properties,
intrinsically related to their functioning as machines.
We have also demonstrated that artificial elastic networks
with similar properties can be constructed
by evolutionary optimization methods.
To verify these theoretical predictions, special single-molecule experiments
monitoring conformational relaxation after arbitrary initial deformations
can be performed.
An example of an elastic machine-like network powered by ligand binding
is presented.
Using such designed networks,
fundamentals of molecular machine operation,
including energetic aspects, the role of thermal noise and
hydrodynamic interactions, can be discussed \cite{Julicher}.
Comparing the behavior of such designed networks with that of
real molecular machines,
better understanding of what is general and what is specific
for a particular protein can be gained.
Moreover, our analysis provides a systematic approach for
the design of proteins with prescribed (programmed) robust
conformational motions.
Not only proteins, but also atomic clusters can be described
by elastic network models \cite{Piazza}.
Similar methods can further be used for engineering of
non-protein machine-like nanodevices (see \cite{Kinbara}).

\appendix
\section{Methods}

The elastic networks for protein structures in Figs. 1 and 2 are
constructed from the structural data of F$_{1}$-ATPase in
the ATP-analog state (Protein Data Bank ID: 1H8H) chain E \cite{1H8H} and
myosin in the ATP-analog state (PDB ID: 1KK8) chain A \cite{1KK8} with
a cutoff distance $l_{0}=10$ {\AA}.
We place a node at the position of each $\alpha$-carbon
atom, and connect nodes lying within the cutoff distance by a link.
Stiffness constants of all links and friction coefficients for all particles
are equal.

To generate random networks, a chain of $N$ nodes is taken and folded
randomly in the three-dimensional space. Each next node is positioned at
random, in such a way that its distance $l$ from the preceding node
satisfies the condition $l_{\min} \leq l \leq l_{\max}$.
Moreover, the next node should be separated by at least
the distance $l_{\min}$ from all previous nodes.
Examining the folded chain, all pairs of nodes ($i$,$j$)
with distances less than $l_{0}$ are noticed ($l_{0} > l_{\max}$) and
connected by elastic strings (hence, not only all neighbors in the string
get connected, but also those nodes which come by chance close one to
another in the folded conformation).

The iterative optimization process consists of a sequence of structural
mutations followed by selection. To perform a mutation, a node is taken at
random and its equilibrium position is changed. The new equilibrium position
is chosen with equal probability within a sphere of radius $d$ around the
old equilibrium position. To preserve the backbone chain, we additionally
require that, after a mutation, the distances of the node from its both
neighbours in the chain lie within an interval from $l_{\min}$ to $l_{\max}$;
other pair distances should not be shorter than $l_{\min}$ either. The new
graph of connections after a mutation is constructed by examining the pair
distances between all nodes and linking the nodes separated at equilibrium
by a distance shorter than $l_{0}$.

Sometimes, networks allowing internal free rotations (and, thus, additional
zero eigenvalues) may be obtained. To distinguish zero eigenvalues
numerically, we adopt a numerical cutoff $\delta\lambda = 10^{-12}$ and
assume the eigenvalues less than $\delta\lambda$ to be zero. For the
networks without internal rotation modes, that is, if the number $N_{nz}$ of
nonzero modes is equal to $3N-6$, we proceed in each iteration step as
described in the main text. When such modes are present, a mutation is
always accepted when it decreases the number of zero eigenvalues
($N_{zm}^{\prime} < N_{zm}$) and accepted with the probability
$P = \exp \left[ -\left( N_{zm}^{\prime}-N_{zm} \right) / \theta \right]$ otherwise.
In this study, we consider networks with $N=64$ nodes. The parameter values
$d=l_{\min}=3.4$, $l_{\max}=4.2$, $l_{0}=8.0$ and $\theta = 0.1$ are used.

Most of the networks after optimization ($97.3\%$) had a large gap $g>3$,
while such gaps were rare in the initial networks ($1.9\%$).
Here, only networks without internal rotations were counted
($1511$ random networks out of $30000$ trials
and $2346$ selected networks out of $2500$ trials).

In the ``reverse'' evolution used to generate special networks with smaller
gaps (such as shown in Fig. 5c), the same evolution algorithm is employed
with the replacement of $g$ by $-g$;
the effective temperature is $\theta = 0.01$ in these simulations.

Around the equilibrium conformation, relative changes $p_{ij}^{(\alpha)}$ in
the distances $u_{ij}$ between nodes $i$ and $j$ in a relaxation
mode $\alpha$ are calculated as
$p_{ij}^{(\alpha)} = \partial{u_{ij}} / \partial{X_{\alpha}} = \mathbf{e}_{ij}^{(\alpha)} \cdot \mathbf{u}_{ij}^{(0)}/u_{ij}^{(0)}$,
where $\mathbf{e}_{ij}^{(\alpha)} = \mathbf{e}_{i}^{(\alpha)}-\mathbf{e}_{j}^{(\alpha)}$, $\mathbf{u}_{ij}^{(0)} = \mathbf{R}_{i}^{(0)}-\mathbf{R}_{j}^{(0)}$ and $u_{ij}^{(0)} = \left\vert \mathbf{u}_{ij}^{(0)} \right\vert$ (see also Eq. (\ref{nmdecay})).
Here, the eigenvectors are normalized as $\sum_{i} \left\vert \mathbf{e}_{i}^{(\alpha)} \right\vert^{2} = 1$.
In Supplementary Fig. 3, statistical distributions of relative changes
$p_{ij}^{(1)}$ in the slowest relaxation mode ($\alpha = 1$)
in ensembles of elastic networks are shown.

For trajectory visualizations, three labels ($1$,$2$,$3$) are attached to a
network and distances $u_{12}$, $u_{13}$ and $u_{23}$ between them are
recorded. The labels are chosen in such a way that $\left\vert p_{12}^{(1)} \right\vert$, the
relative distance change in the slowest relaxation mode, is maximal between
the nodes labeled as $1$ and $2$; then the third node is chosen for which
$\left\vert p_{13}^{(2)} \right\vert$, in the second slowest relaxation mode, is maximal between the
nodes labeled as $1$ and $3$ (there are two choices and we have selected one
of them).

To prepare initial conditions when relaxation patterns are considered, we
apply randomly chosen static forces $\mathbf{F}_{i,static}$ and wait for a
certain time $\tau_{s}$. These random forces are chosen in such a way that
their total magnitude
$F_{s} = \left[ \sum_{i=1}^{N} \left\vert \mathbf{F}_{i,static} \right\vert^{2} \right] ^{1/2}$ is fixed.
Then, we remove the
static forces and record the trajectory. For each network, 100 trajectories
from different initial conditions are shown.
In Fig. 1, $F_{s}=10$, $\tau_{s}=3\times 10^{4}$;
in Fig. 2, $F_{s}=0.1$, $\tau_{s}=10^{5}$;
and in Figs. 4 and 5, Supplementary Figs. 1 and 2, $F_{s}=0.1$ and $\tau_{s}=10^{6}$.

In Fig. 6 and in Supplementary Video 1, we initially place a ligand in the
center of mass of the three ligand-binding nodes and introduce additional
links to these nodes with the natural length of $1.7$. At the initial
moment, the links are longer than their natural lengths (i.e., they are
stretched) and thus attractive forces between the ligand and the binding
nodes are acting. To include fluctuations, random time-dependent forces
$\mathbf{f}_{i}(t)$ of intensity $\sigma$ have been added to the equations of
motion of all particles, i.e.,
$\overset{.}{\mathbf{R}}_{i} = \mathbf{F}_{i,elastic} + \mathbf{f}_{i}(t)$ with
$\langle \mathbf{f}_{i}(t) \rangle = \mathbf{0}$ and
$\langle \mathbf{f}_{i}(t) \cdot \mathbf{f}_{j}(t^{\prime}) \rangle = 6\sigma \delta_{ij} \delta(t-t^{\prime})$.
We have $\sigma=10^{-3}$ (upper panel in Fig. 6 and Supplementary Video 1)
and $3\times10^{-3}$ (lower left panel in Fig. 6).
At $t=20000$ (which is long enough for the
network to relax to the steady state with a ligand, snapshot D), we remove
the ligand and the additional links. In the ligand-free state, we have
assumed that the network binds with a ligand again at a constant rate
(probability per unit time) $\nu$ even in nonequilibrium network
configurations, provided that all distances $u_{ij}$ between the three
ligand-binding nodes satisfy the conditions
$\left\vert \Delta u_{ij} \right\vert = \left\vert u_{ij}-u_{ij}^{(0)} \right\vert \leq \epsilon_{l}$,
i.e., the conformation of the ligand-binding site is close enough to the
initial one. Then, a ligand is placed again in the center of mass of the
ligand-binding nodes and the next cycle starts. The parameter values
$\nu=10^{-6}$ and $\epsilon_{l}=0.01$ are used. Coordinates of the nodes in
the constructed example of the machine-like network and positions of the
three binding nodes are given in Supporting Information Data 1.

\begin{acknowledgments}
Financial support of Japan Society for the Promotion of Science
through the fellowship for research abroad (H17) for Y. T. is
gratefully acknowledged.
% The authors declare that they have no competing financial interests.
\end{acknowledgments}

%--- references ---

\begin{figure*}
\begin{center}
\includegraphics[width=134.4mm]{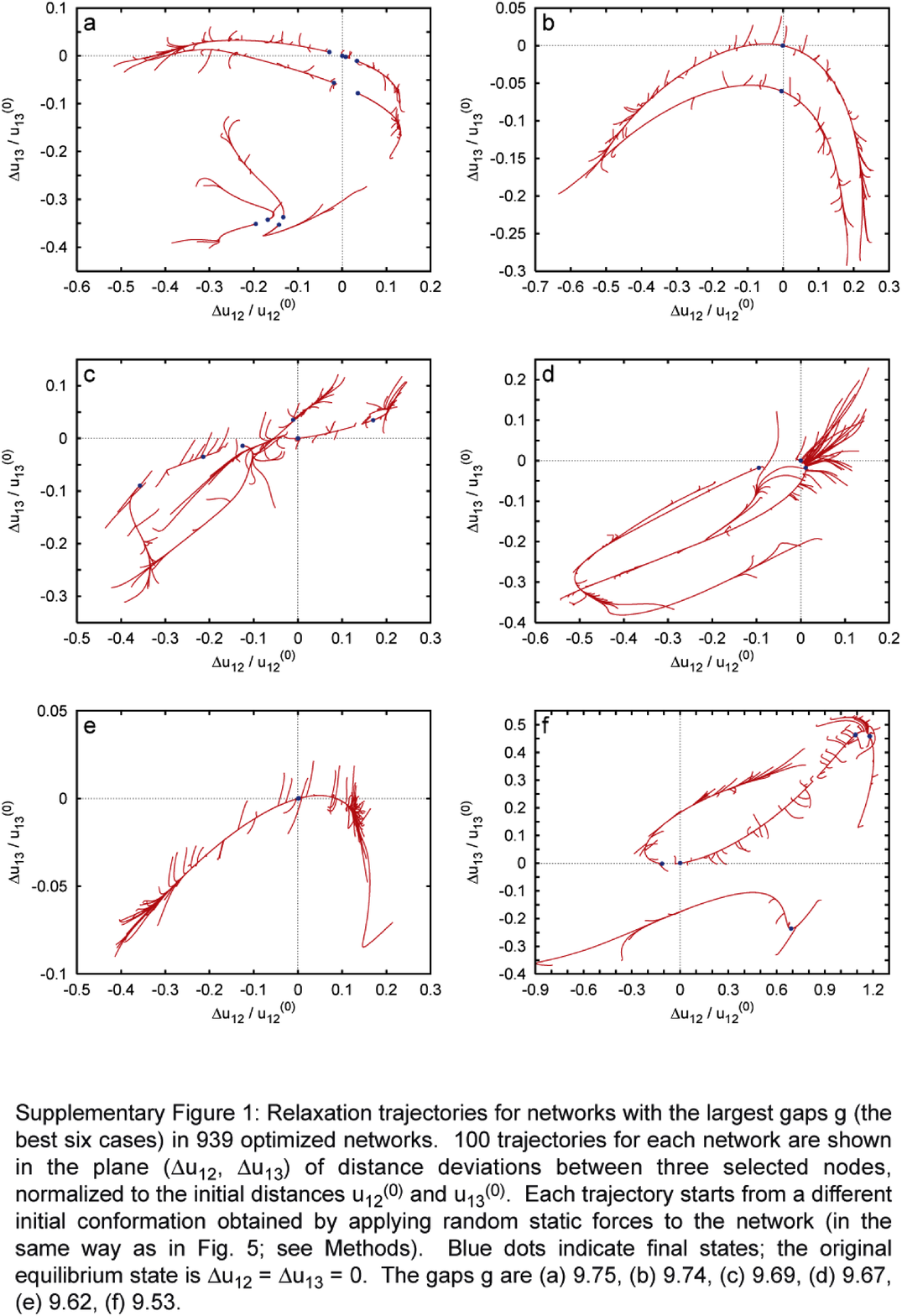}
\end{center}
\end{figure*}

\begin{figure*}
\begin{center}
\includegraphics[width=135.06667mm]{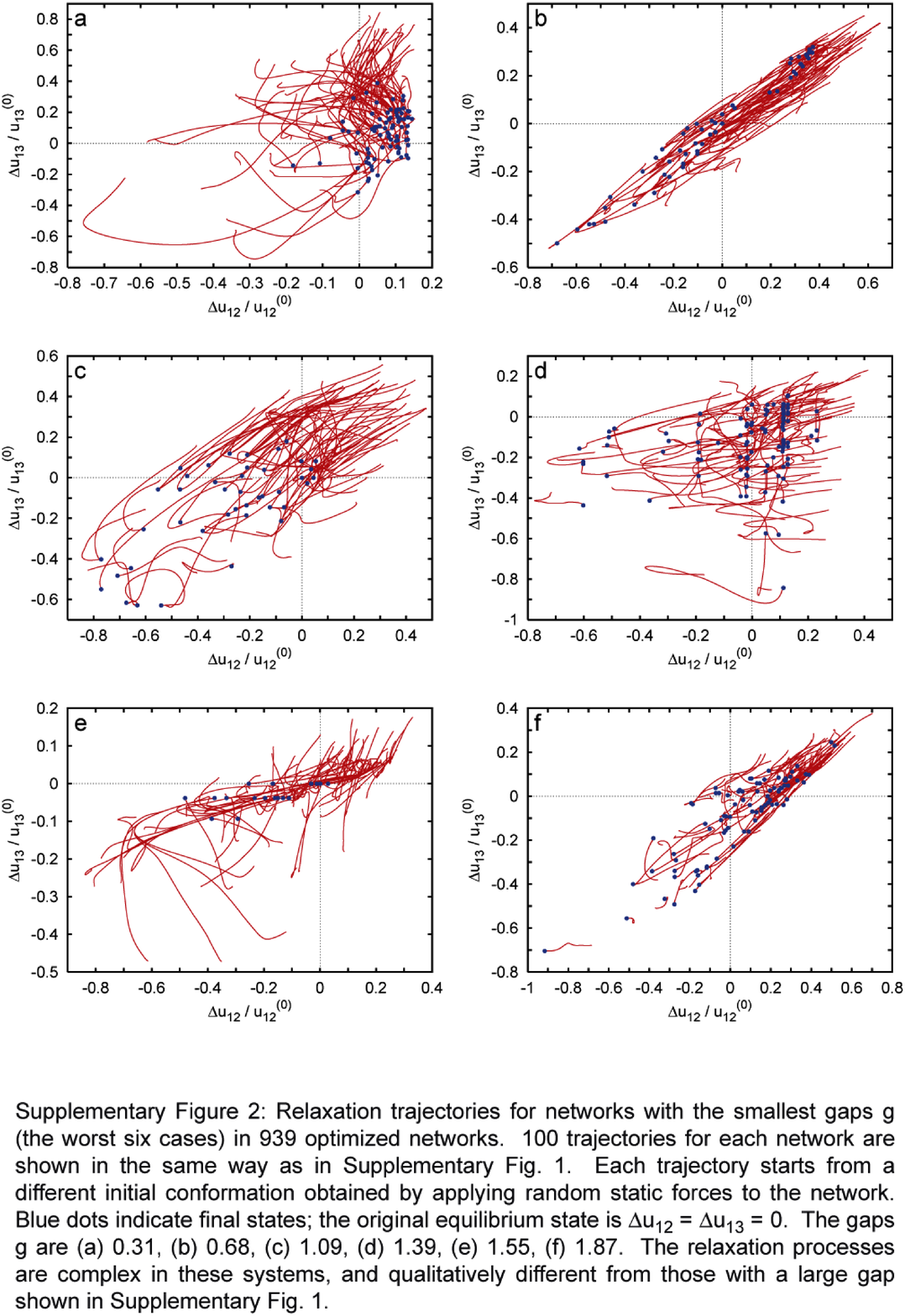}
\end{center}
\end{figure*}

\begin{figure*}
\begin{center}
\includegraphics[width=135.33333mm]{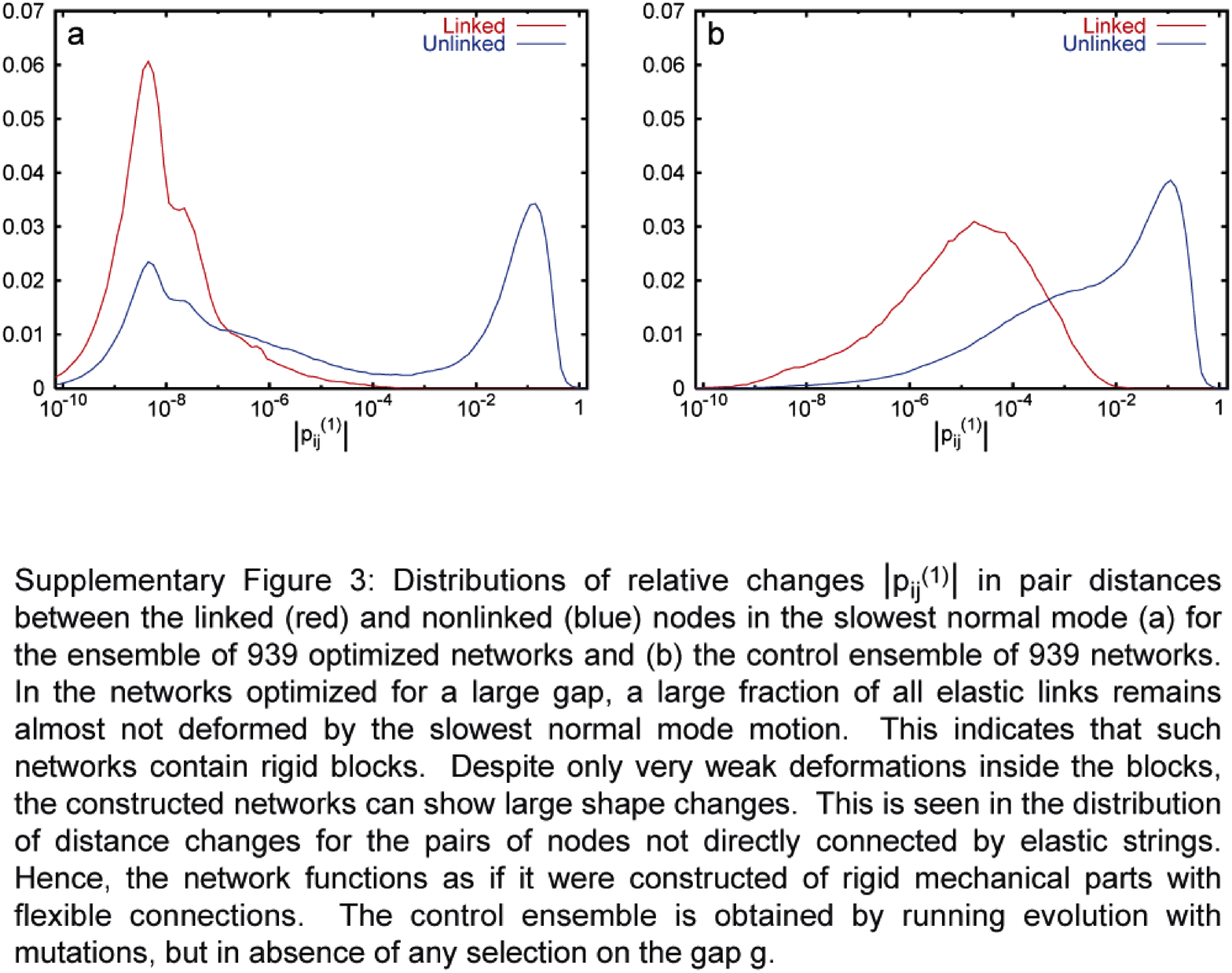}
\end{center}
\end{figure*}

\end{document}